\documentclass[reprint, nofootinbib, prd, twocolumn, superscriptaddress,
longbibliography, preprintnumbers]{revtex4-2}

\usepackage{relsize}

\usepackage[acronyms, nohypertypes={acronym}, nopostdot, style=super, nonumberlist, toc]{glossaries}
\glsenableentrycount
\setacronymstyle{long-sc-short}

\newcommand\ac[1]{\gls{#1}}

\def\acresetall{\glsresetall[acronym]}

\newacronym{WF}{wf}{Wilson-Fisher}
\newacronym{AF}{af}{asymptotically free}
\newacronym{RG}{rg}{renormalization group}
\newacronym{WZW}{wzw}{Wess-Zumino-Witten}
\newacronym[longplural={conformal field theories}]{CFT}{cft}{conformal field theory}
\newacronym[longplural={lattice field theories}]{LFT}{lft}{lattice field theory}
\newacronym[longplural={effective field theories}]{EFT}{eft}{effective field theory}
\newacronym[longplural={quantum field theories}]{QFT}{qft}{quantum field theory}
\newacronym{LEC}{lec}{low-energy constant}
\newacronym{QCD}{qcd}{quantum chromodynamics}
\newacronym{MC}{mc}{Monte Carlo}
\newacronym{IR}{ir}{infrared}
\newacronym{UV}{uv}{ultraviolet}
\newacronym{SNR}{snr}{signal-to-noise ratio}
\newacronym{NLSM}{nl$\sigma$m}{nonlinear sigma model}
\newacronym{PCM}{pcm}{principal chiral model}
\newacronym{CSA}{csa}{Cartan subalgebra}
\newacronym{SSB}{ssb}{spontaneous symmetry breaking}
\newacronym{DOF}{dof}{degrees of freedom}
\newacronym{DMRG}{dmrg}{densiy matrix renormalization group}
\newacronym{YM}{ym}{Yang-Mills}
\newacronym{QLM}{qlm}{quantum link model}
\newacronym{KG}{kg}{Kogut-Susskind}
\newacronym{SPT}{spt}{symmetry protected topological}
\newacronym{GW}{gw}{Ginsparg-Wilson}
\newacronym{FK}{fk}{Fidkowski-Kitaev}
\newacronym{CS}{cs}{Chern-Simons}
\newacronym{APS}{aps}{Atiyah-Patodi-Singer}
\newacronym{PV}{pv}{Pauli-Villars}
\newacronym{PBC}{pbc}{periodic boundary conditions}
\newacronym{OBC}{obc}{open boundary conditions}
\newacronym{ABC}{abc}{antiperiodic boundary conditions}
\newacronym{KD}{kd}{Kähler-Dirac}
\newacronym{SMG}{smg}{symmetric mass generation}
\newacronym{GB}{cgb}{Chern-Gauss-Bonnet}
\newacronym{WY}{wy}{Witten-Yonekura}
\newacronym{RKD}{rkd}{reduced Kähler-Dirac}
\newacronym{AS}{as}{Atiyah-Singer}
\newacronym{BC}{bc}{boundary conditions}

\usepackage{amsmath,amssymb,amsbsy,amsfonts}
\usepackage{mathtools}
\usepackage{bm}
\usepackage{microtype}
\usepackage{dutchcal}

\usepackage{enumitem}
\usepackage{graphicx}
\usepackage{amsfonts}
\usepackage{tocvsec2}
\usepackage{slashed}
\usepackage{cancel}
\usepackage{comment}
\usepackage{ulem}
\usepackage{minibox}
\usepackage{orcidlink}

\usepackage{graphics}
\graphicspath{{./fig/}}

\usepackage{hyperref}
\hypersetup{
    unicode,
    pdfprintscaling=None,
    colorlinks,
    breaklinks
}
\hypersetup{
  colorlinks=true,   % false: boxed links; true: colored links
  linkcolor=cyan,    % color of internal links (box color = linkbordercolor)
  citecolor=blue,    % color of links to bibliography
  filecolor=magenta, % color of file links
  urlcolor=blue      % color of external links
}

\usepackage[capitalise]{cleveref}
\crefname{section}{Sec.}{Secs.}

\newcommand\<{\langle}
\renewcommand\>{\rangle}

\newcommand\xdagger{\vphantom{\dagger}}

\newcommand\del\partial

\newcommand\cD{\mathcal{D}}

\newcommand\Id{\mathbb{I}}

\usepackage{makecell}

\newcommand\beq{\begin{eqnarray}}
\newcommand\eeq{\end{eqnarray}}

\newcommand{\mybar}[1]{\kern 0.6pt\overline{\kern -0.6pt#1\kern -0.6pt}\kern 0.6pt}

\def\U\Omega{U(1)_{\Omega}}

\newcommand{\Psibar}{\bar\Psi}

\DeclareMathOperator{\tr}{tr}

\newcommand{\Phibar}{\bar\Phi}

\newcommand\GammaChi{\Gamma}
\newcommand\GammaGB{\Gamma_\chi}
\newcommand\GammaH{\Gamma_\sigma}

\newcommand{\cK}{\mathcal{K}}

\newcommand\RR{\mathbb{R}}

\newcommand\Spin{\mathrm{Spin}}

\newcommand\KD{{\text{KD}}}

\newcommand{\gammabar}{\Bar{\gamma}}
\newcommand\Ky{\cK_{Y}}
\newcommand\Kx{\cK_{X}}
\newcommand\QA{Q}
\newcommand\hsp{\mathcal{H}}
\newcommand{\ind}{I}

\usepackage{totcount}
\newtotcounter{commentcount}
\regtotcounter{commentcount}

\begin{document}

\preprint{FERMILAB-PUB-24-0262-T}

\title{Chiral symmetry and Atiyah-Patodi-Singer index theorem for staggered fermions}

\author{Mendel Nguyen\,\orcidlink{0000-0002-7976-426X}}
\email{mendelnguyen@gmail.com}
\affiliation{Department of Physics, North Carolina State University, Raleigh, North Carolina 27607, USA}

\author{Hersh Singh\,\orcidlink{0000-0002-2002-6959}}
\email{hershs@fnal.gov}
\affiliation{Fermi National Accelerator Laboratory, Batavia, Illinois 60510, USA}

\begin{abstract}
  We consider the \ac{APS} index theorem corresponding to the chiral symmetry of a continuum formulation of staggered fermions called K\"ahler-Dirac fermions, which have been recently investigated as an ingredient in lattice constructions of chiral gauge theories.
  We point out that there are two notions of chiral symmetry for K\"ahler-Dirac fermions, both having a mixed perturbative anomaly with gravity leading to index theorems on closed manifolds.
  By formulating these theories on a manifold with boundary, we find the \ac{APS} index theorems corresponding to each of these symmetries, necessary for a complete picture of anomaly inflow, using a recently discovered physics-motivated proof.
  We comment on a fundamental difference between the nature of these two symmetries by showing that a sensible local, symmetric boundary condition only exists for one of the two symmetries.
  This sheds light on how these symmetries
  behave under lattice discretization, and in particular on their use for recent \ac{SMG} proposals.
\end{abstract}

\maketitle

\acresetall

\section{Introduction}

A lattice definition of chiral gauge theories has been a major challenge in nonperturbative quantum field theory ever since the problem was put in sharp focus by Nielsen and Ninomiya
\cite{Nielsen:1980rz}.
Despite remarkable progress in constructing doubler-free lattice formulations of Dirac fermions with good chiral properties exploiting the \ac{GW} relation \cite{Ginsparg:1981bj, Kaplan:1992bt, Narayanan:1993ss, shamir_chiral_1993, Neuberger:1997fp}, and even in constructing abelian chiral theories using this approach \cite{luscher2001chiral, luscher1999abelian}, the general problem for non-Abelian chiral theories remains open \cite{luscher2001chiral, kikukawa_domain_2002a, kaplan_chiral_2012}.

Recently, there has been renewed intensity in efforts towards solving this problem
\cite{aoki_lattice_2024a, aoki_lattice_2024b, berkowitz_exact_2023, catterall_chiral_2021, catterall_lattice_2023, kaplan_chiral_2024, kaplan_weyl_2024, pedersen_reformulation_2023, wang_nonperturbative_2022, wang_solution_2019, wang_symmetric_2022a, zeng_symmetric_2022, golterman_conserved_2024, golterman_propagator_2024, clancy_generalized_2024}, largely owing to a refined understanding of the correspondence between topological phases in the bulk and 't~Hooft anomalies at the boundary.
The bulk--boundary correspondence states that 't~Hooft anomalies of $d$-dimensional theories are in one-to-one correspondence with $(d+1)$-dimensional \ac{SPT} phases.
This can be thought of as a broad generalization of the Callan-Harvey \cite{Callan:1984sa} anomaly inflow mechanism, incorporating more general global anomalies \cite{witten_anomaly_2020, witten_fermion_2016}.

In all recent approaches to the chiral gauge theory problem, the bulk--boundary correspondence plays an important role.
As emphasized by Witten \cite{witten_fermion_2016}, a key ingredient in this picture is the \ac{APS} index theorem \cite{atiyah_spectral_1975}, a generalization of the \ac{AS} index theorem to manifolds with boundary.
On closed manifolds, the usual \ac{AS} index theorem relates the index of a Dirac operator to a topological quantity constructed in terms of gauge fields.
However, on a manifold with boundary, this gets modified to include an extra term from the boundary, called the $\eta$ invariant.
It is in fact possible to systematically understand fermionic \ac{SPT} phases and their anomaly inflow in terms of the $\eta$-invariant
\cite{kapustin_fermionic_2015, witten_anomaly_2020, witten_fermion_2016},
incorporating general ``Dai-Freed'' anomalies recently discusssed in the literature
\cite{dai1994eta, yonekura_daifreed_2016, garcia-etxebarria_daifreed_2019}.
It is therefore desirable to understand the various approaches being pursued for constructing chiral gauge theories in this general picture.

In one of these approaches, known as \ac{SMG}
\cite{eichten1986chiral, wang_symmetric_2022a}, one starts with a chiral theory and adds mirror fermions to construct a vector-like theory, hoping to eventually gap out the mirror fermions by designing suitable interactions.
For this scenario to occur, a necessary (though not sufficient) condition is the cancellation of 't~Hooft anomalies.
Over the years, this has been studied in various guises
\cite{ayyar_fermion_2016, ayyar_generating_2017, ayyar_origin_2016, catterall_chiral_2021, catterall_topology_2018, razamat_gapped_2021, you_interacting_2015, butt_anomalies_2021, catterall_chiral_2021, catterall_lattice_2023, catterall_topological_2018, butt_invariant_2018, creutz_lattice_1997, kikukawa_why_2019, poppitz_lattice_2007}.
The main difficulty comes from the fact that,
although \ac{SMG} is well-understood in two dimensions
\cite{zeng_symmetric_2022, wang_solution_2019, wang_nonperturbative_2022, wang_symmetric_2022a}, designing the right interactions in higher dimensions is in general a delicate task.\footnote{See also Ref.~\cite{golterman_propagator_2024} for another recent criticism.}

An elegant approach to constructing \ac{SMG} interactions has been through the use of \ac{KD} fermions \cite{kaehler_innere_1962}.
The \ac{KD} equation provides an alternative to the Dirac equation and admits a natural lattice discretization which preserves many properties from the continuum 
\cite{banks_geometric_1982a, becher_dirackahler_1982, rabin_homology_1982}.
It was recognized early on as providing a geometric formulation of staggered fermions \cite{susskind_lattice_1977, sharatchandra_susskind_1981, rabin_homology_1982}.
\ac{KD} fermions also appear naturally in lattice supersymmetry \cite{catterall_exact_2009}, and
recently have been investigated in the context of \ac{SMG}
of staggered fermions \cite{catterall_topology_2018, ayyar_origin_2016, ayyar_generating_2017, ayyar_fermion_2016} and construction of chiral lattice gauge theories
\cite{catterall_chiral_2021, catterall_chiral_2021, catterall_hooft_2022}.

The fact that two \ac{KD} fermions have the right number of Majorana/Weyl flavors in flat spacetime required for \ac{SMG} to occur \cite{guo_symmetric_2023, catterall_hooft_2022} can be explained by showing the cancellation of 't~Hooft anomalies exactly on the lattice \cite{catterall_hooft_2022}, or by mapping them to bosonic \ac{SPT} phases \cite{guo_symmetric_2023}, providing evidence for the usefulness of studying the bulk--boundary correspondence for \ac{KD} fermions.
Given the relevance of the \ac{APS} index theorem for formulating the bulk--boundary correspondence for fermions, it is therefore important to understand how the \ac{APS} index theorem manifests for \ac{KD} fermions.

This brings us to the point of this work. 
How does the \ac{APS} index theorem work for the chiral symmetry of \ac{KD} fermions? 
And given that \ac{KD} fermions have a natural lattice discretization which preserves (some of) the anomalies, does the \ac{APS} index theorem have a natural lattice version as well?
Recently, Kobayashi and Yonekura \cite{kobayashi_atiyahpatodisinger_2021} have adapted Fujikawa's method \cite{fujikawa1979path} to manifolds with boundary in a way which naturally leads to the \ac{APS} index theorem.
In particular, they use a physical interpretation of the \ac{APS} boundary conditions and the $\eta$ invariant, which makes the proof quite transparent.
In this work, we adapt their strategy to understand the \ac{APS} theorem for \ac{KD} fermions, clarifying some important differences from the case of Dirac fermions considered by them.

Our set up is the following. 
We consider \ac{KD} fermions in even-dimensional (Euclidean) spacetime dimensions coupled to gravity.
For any global symmetry of the theory in flat spacetime, we can now ask whether this symmetry survives turning on a nontrivial background gravitational field -- that is, whether this symmetry is present on arbitrary curved manifolds.
For us, the global symmetry of interest is a type of chiral symmetry.
So the question is whether a theory of \ac{KD} fermions in arbitrary curved space has a chiral symmetry.
If not, we say that there is a mixed 't~Hooft anomaly between gravity and chiral symmetry.

Actually, it turns out that \ac{KD} fermions admit two notions of chiral, or chiral-like, symmetry.
One of them survives the natural lattice discretization, and the other one does not \cite{becher_dirackahler_1982, catterall_hooft_2022}.
We will discuss both of these symmetries in this work, and in particular, clarify the differences in their mixed anomalies with gravity and their associated index theorems.
(See also Refs.~\cite{gockeler_axialvector_1983,linhares_axial_1985, shimono_fermions_1990} for older work focusing on the mixed $U(1)$ vector-axial anomaly.)
Even though the two symmetries are analogous on closed manifolds,
the difference between them is particularly salient on manifolds with boundary.
This points to a fundamental difference in the nature of their anomalies and \ac{APS} index theorems.

This article is organized as follows. We review some basic notions about \ac{KD} fermions in \cref{sec:kd}, emphasizing the two different notions of chiral symmetry.
In \cref{sec:anomaly-perturbative}, we recall that both these chiral symmetries have a perturbative 't~Hooft anomaly with gravity, and show that these lead to two well-known index theorems.  
In \cref{sec:aps}, we prove the \ac{APS} index theorem corresponding to both these symmetries.  
Finally, in \cref{sec:boundary-conditions}, we remark on boundary conditions which distinguish the nature of two symmetries and their anomalies in a fundamental way, explaining when the \ac{APS} index theorem does not get a boundary contribution.
In \cref{sec:conclusion}, we summarize this work and present our conclusions.

\section{Kähler-Dirac fermions}
\label{sec:kd}

In this section, we review the continuum formalism for \ac{KD} fermions \cite{becher_dirackahler_1982,banks_geometric_1982a,kaehler_innere_1962}.
Throughout this work, $X$ denotes a $D=(d+1)$-dimensional manifold with Euclidean signature metric $g_{\mu\nu}$.

There are two useful ways of talking about \ac{KD} fields.
The first uses the language of differential forms. (See Appendix~\ref{sec:dforms} for a review.)
Given a collection of complex Grassmann-valued skew-symmetric covariant tensors $\phi_{\mu_1 \ldots \mu_p}$ of rank $p$ for each $p=0,1,\dotsc,D$, we may define a \ac{KD} field by the formal sum
\begin{align}
    \bm\phi = \sum_{p=0}^D \frac{1}{p!} \phi_{\mu_1\ldots\mu_p} dx^{\mu_1} \wedge \ldots \wedge dx^{\mu_p}
\end{align}
In the second way, we simply replace each coordinate differential $dx^\mu$ in the above expression by the curved space gamma matrix $\Gamma^\mu$ (see Appendix~\ref{sec:vielbein}), resulting in the $2^{D/2}\times2^{D/2}$ matrix-valued field
\begin{align}
    \Phi = \sum_{p=0}^D \frac{1}{p!} \phi_{\mu_1\ldots\mu_p} \Gamma^{\mu_1} \ldots \Gamma^{\mu_p}
\end{align}
Both notations are useful.
We use bold lowercase greek letters ($\bm\phi, \bm\psi, \dotsc$) to denote \ac{KD} fields in differential form notation, while we use uppercase greek letters to denote \ac{KD} fields in matrix notation ($\Phi, \Psi, \dotsc$)

In differential form notation, the \ac{KD} operator is given by $\cK = i(d - \delta)$ where $\delta$ is the adjoint of $d$.
Note that it squares to $\cK^2 = d\delta + \delta d$, which is the Laplacian.
In matrix notation, the \ac{KD} operator is given by $\cK = i \Gamma^\mu D_\mu$ where $D_\mu$ is the gravitational covariant derivative.
The \ac{KD} action is given by
\begin{align}
    S_{\KD} &= \int_{X} \bar{\bm\phi}\wedge \star \cK {\bm\phi}
    = \int_X \tr \bigl( \Phibar \cK \Phi\bigr)\sqrt{g}d^Dx
\end{align}
where in the middle expression, it is implied that only top degree forms are selected in the integrand.

The matrix notation makes it clear that in flat space, a \ac{KD} fermion may be interpreted as a multiplet of $2^{D/2}$ flavors of Dirac fermions, the flavors being the columns of $\Phi$. 
Clearly, $S_{\KD}$ has a flavor symmetry under $\Phi \to \Phi F^\dag$ with $F \in SU(2^{D/2})$.

\subsection*{Two chiral symmetries}
\label{sec:chiral-sym}

As mentioned in the introduction, \ac{KD} fermions enjoy two notions of chiral symmetry.
Since in this work we will be comparing these two symmetries, we review here how they may be found. 
(See Ref.~\cite{green_superstring_2012} for a somewhat different discussion.)

By a chirality operator, we mean an operator $\GammaChi$ that squares to the identity and anticommutes with the \ac{KD} operator:
\begin{align}
    &\GammaChi^2 = 1,\\
    & \{\cK,\GammaChi\} = 0.
\end{align}
There are two obvious ways for $\GammaChi$ to anticommute with $\cK$: either it anticommutes with $d$ and $\delta$ individually, or it exchanges $d$ and $\delta$.
Both of these actually lead to definitions of a chiral symmetry.

If we require $\GammaChi$ to anticommute with both $d$ and $\delta$, then we obtain the operator $\GammaChi = \GammaGB$ given by
\begin{align}
    \GammaGB \equiv (-1)^{k}
\end{align}
on $k$-forms.

If we require $\GammaChi$ to exchange $d$ and $\delta$, then we instead obtain the operator $\GammaChi = \GammaH$ given by (see Appendix~\ref{sec:ChiralHsym})
\begin{align}
    \GammaH \equiv i^{\frac12 D(D+1)} 
 (-1)^{\frac12 k(k+1)} \star
\end{align}
on $k$-forms.
Note that in dimensions $D$ a multiple of $4$, we have $\GammaH=\star$ on $\frac12 D$-forms.

In the matrix notation, the two chirality operators are given very simply as
\begin{align}
    &\GammaGB(\Psi) = \gammabar \Psi \gammabar, \\
    &\GammaH(\Psi) = \gammabar \Psi,
              \label{eq:two-chiral-sym}
\end{align}
where $\gammabar$ is the flat space ``$\gamma^{5}$'' matrix in even dimensions:
\begin{align}
	\gammabar = i^{D(D-1)/2} \gamma_{1} \gamma_{2} \cdots \gamma_{D}.
\end{align}
We see that the symmetry generated by $\GammaH$ is essentially the obvious generalization of the chiral symmetry of Dirac fermions, whereas the symmetry generated by $\GammaGB$ also involves a transformation of the flavor indices.
For this reason, the latter is sometimes referred to as the twisted chiral symmetry, and we adopt this terminology here.
We shall refer to the chiral symmetry generated by $\GammaH$ as the untwisted chiral symmetry. 

Using either chirality operator, it is possible to halve the number of degrees of freedom by projecting onto one of the $\Gamma = \pm1$ eigenspaces. 
It is more common, however, to use the twisted chirality operator $\GammaGB$, since this survives discretization while the untwisted chirality operator $\GammaH$ does not.
\ac{KD} fermions of definite twisted chirality are known as reduced \ac{KD} fermions, and upon discretization are completely equivalent to reduced staggered fermions. 

\section{Perturbative chiral anomaly}
\label{sec:anomaly-perturbative}

In analogy with the chiral symmetry for Dirac fermions in even dimensions, the two chiral symmetries discussed in the previous section are anomalous.
To be more precise, there is a perturbative 't Hooft anomaly involving the chiral symmetry $\GammaChi$ and gravity, for both the twisted and untwisted chiral symmetries.

It has been shown that \ac{KD} fermions have a perturbative mixed anomaly involving  twisted chiral symmetry and gravity in Refs.~\cite{catterall_topological_2018, butt_anomalies_2021,catterall_hooft_2022,catterall_induced_2022}.
The index theorem associated with this anomaly is the \ac{GB} theorem.
However, the more standard notion of chiral symmetry corresponds to the untwisted chiral symmetry. This symmetry also participates in a perturbative mixed anomaly with gravity in $D=4n$ dimensions. As we will show in this section, the corresponding index theorem is the Hirzebruch signature theorem.

On manifolds without boundary, the mixed chiral--gravitational anomaly can be found by examining the change in the path integral measure, following a standard Fujikawa type analysis.
Since the analysis is identical for both chiral symmetries,  we proceed in a general way, using the notation $\GammaChi$ for either symmetry.

As usual, to define the path integral, we expand the fields in an appropriate set of normal modes, and take the path integral to mean the Grassmann integral over the expansion coefficients.
Here, the normal modes will be taken to be eigenfunctions of the \ac{KD} operator $\cK = i (d - \delta)$.

As is familiar from related discussions of fermion anomalies, the fact that $\GammaChi$ anticommutes with the \ac{KD} operator implies a complete pairing of nonzero modes.
Specifically, if $\xi_n$ is an eigenfunction of $\cK$ with nonzero eigenvalue $\lambda_n$, then $\GammaChi \xi_n$ is an independent eigenfunction with eigenvalue $-\lambda_n$.
Since on the zero mode subspace, $\GammaChi$ commutes with $\cK$, we may take the zero modes $\zeta_k$ to have definite chirality $\GammaChi=\pm 1$.
If we write the expansion of $\bm \phi$ as
\begin{align}
    \bm \phi = \sum_n (\xi_n a_n + \GammaChi \xi_n b_n) + \sum_k \zeta_k c_k,
\end{align}
where the $a_n,b_n,c_k$ are Grassmann-valued coefficients, then the path integral measure is given by
\begin{align}
    [d \bm \phi] = \prod_n da_n db_n \prod_k dc_k.
\end{align}
We now examine the Jacobian for the change of variables ${\bm \phi} \to  e^{i \alpha \GammaChi} {\bm \phi}$.
A standard argument shows that each pair of nonzero modes contributes a factor of $1$ to the Jacobian determinant, which is therefore entirely determined by the zero modes.
The zero mode expansion coefficients transform simply as $c_k \to  e^{\pm i \alpha} c_k$, the sign $\pm$ being chirality of $\zeta_k$.
It follows that the path integral measure is transformed as
\begin{align}
  [d {\bm \phi}] &\to  [d {\bm \phi}] \exp [-i \alpha (N_+ - N_-)] ,
    \label{eq:jacobian-anomaly}
\end{align}
where $N_\pm$ is the number of zero modes with $\GammaChi = \pm 1$, and $I = \tr \Gamma \cK =  N_{+} - N_{-}$ is the index of the \ac{KD} operator corresponding to the $\Gamma$ symmetry.
The transformation of the $\bar {\bm \phi}$ field as $\bar {\bm \phi} \to \bar {\bm \phi} e^{{i \alpha \Gamma }} $ identically transforms the measure $[d \bar {\bm \phi}]$,  resulting in the full Jacobian \begin{align}
  [d \bar {\bm \phi} ] [d {\bm \phi}] &\to 
                      [d \bar {\bm \phi} ] [d {\bm \phi}]
                      \exp [- 2 i \alpha \tr (\Gamma \cK) ].
\end{align}

To relate the index to more familiar topological invariants, it is useful to go back to the language of differential forms and recall a few definitions. If $\Delta = \cK^{2}$ is the Laplacian on the manifold $X$, then the zero modes of $\Delta$ are called harmonic forms.
Let $b_{p}$ ($p=0,1,\dotsc,D$) be the number of linearly independent degree-$p$ harmonic forms, also called the Betti numbers of the manifold. 
Since $[\Gamma, \Delta] = 0$, we can take the harmonic forms to have definite chirality for either definition of the chiral symmetry.

\subsection{The Euler-characteristic anomaly}

Let us first consider the twisted chiral symmetry \cite{catterall_topological_2018, butt_anomalies_2021,catterall_hooft_2022,catterall_induced_2022}. In this case, the number of positive (negative) chirality zero modes are just the even (odd) degree harmonic forms.  The fact that the index is the Euler characteristic can be seen from the definition of Euler characteristic as an alternating sum of Betti numbers $\chi = \sum_{p} (-1)^{p} b_{p} = N_{+} - N_{-}$.

To obtain the other side of the index theorem, relating the Euler characteristic to an integral of a local density of the curvature form $R$, one computes the trace in the Jacobian in \cref{eq:jacobian-anomaly} using a gauge-invariant heat-kernel regularization \cite{fujikawa1979path}.  The computations result in the Euler density, as confirmed in Ref.~\cite{catterall_induced_2022}, leading to the well-known \ac{GB} theorem
\begin{align}
   \chi = \int_{X} e(R)
\end{align}
where $e(R)$ is the Euler density. For example, in $D=4$ dimensions, $e(R) \propto \varepsilon_{mnrs} R_{mn} \wedge R_{rs}$.

\subsection{The signature anomaly}

Now, let us consider the untwisted chiral symmetry.
In this case, it turns out that the quantity $N_+ - N_-$ is the signature $\sigma$ of the manifold, which is non-vanishing only for dimensions $D$ a multiple of four.
Writing $b_\pm$ for the number of degree-$D/2$ harmonic forms $\zeta$ with $\star \zeta = \pm \zeta$ (that is, self-dual harmonic forms), the signature $\sigma(X)$ of a manifold $X$ is a topological invariant defined to be
\begin{equation}
    \sigma = b_+ - b_-.
\end{equation}
The Hirzebruch signature theorem gives the signature as a integral over the manifold
\begin{equation}
    \sigma = \int_X L(R)
\end{equation}
where $L(R)$, known as the Hirzebruch $L$ polynomial, is a polynomial in the Pontrjagin classes.
For example, in $D=4$, we have $L(R) \propto \tr R \wedge R $.

To see that $N_+ - N_-$ is indeed the signature, we first note the set of zero-modes of the \ac{KD} operator $\cK$ is precisely the set of harmonic forms on $X$.
Indeed, we have
\begin{align}
	(\zeta , \Delta \zeta) = (\zeta, \cK^\dag \cK \zeta) = (\cK \zeta, \cK \zeta)
\end{align}
so that $\cK \zeta = 0$ if and only if $\Delta \zeta = 0$.
So $N_\pm$ is the number of independent harmonic forms with $\GammaH = \pm 1$.

Denote the subspace containing only degree $p$ and $(D-p)$ harmonic forms as $H_{p}$, with $p < D/2$.
Let $\zeta_1,\ldots,\zeta_N$ be a complete set of independent harmonic forms of degree $p$.
Then by Poincar\'{e} duality, $\GammaH\zeta_1,\ldots,\GammaH\zeta_N$ is a complete set of independent harmonic forms of degree $(D-p)$.
Now, the linear combinations $(1 \pm \GammaH) \zeta_{i}$ are also harmonic, have a definite chirality $\GammaH = \pm1 $, and span the subspace $H_{p}$. Therefore, there are as many positive chirality modes in $H_{p}$ as there are negative chirality modes. Since this is true for all $p < D/2$, the contributions from all forms with degree $p \neq D/2$ cancel when computing the index $N_{+} - N_{-}$ .
The only contributions to the index can come from forms of degree $D/2$, where we do not have such a pairing between positive and negative chirality modes.  But $\GammaH = \star$ on middle forms, so $N_{+} - N_{{-}} = b_{+} - b_{-}$, which is the signature.

We note that for reduced \ac{KD}, the measure $[d\Bar{{\bm \phi}}]$ will not transform, as it has no mode with even form degree and therefore cannot contribute to the signature.
So reduced \ac{KD} have half the anomaly.

\section{APS theorem for Kahler-Dirac Fermions}
\label{sec:aps}

On a closed manifold $X$, the \ac{AS} index theorem equates the index $I$ of a differential operator $\cD_X$ to the integral over $X$ of a local density $G$ involving gauge fields: $I = \int_X G$.
The \ac{APS} index theorem is a far-reaching generalization of the \ac{AS} theorem for manifolds with boundary.
For a specific choice of boundary condition, the \ac{APS} theorem states that the \ac{AS} index formula is corrected by a boundary contribution:
\begin{align}
    I_{\textsc{aps}} = \int_X G + \frac{1}{2} \eta_{Y}.
\end{align}
The boundary contribution $\eta_{Y}$ is a spectral invariant of a differential operator on the boundary.

Since the more familiar \ac{AS} index theorem may be derived essentially via Fujikawa's analysis of perturbative fermion anomalies, it is natural to ask if a similar thing can be done for the \ac{APS} index theorem.
Kobayashi and Yonekura \cite{kobayashi_atiyahpatodisinger_2021} have shown that it can for the case of Dirac fermions.
Following their methods, we show in this section that an \ac{APS} index theorem can also be derived for \ac{KD} fermions with respect to both notions of the chiral symmetry.

The setup of the theorem is as follows.
We take $X$ to be a Euclidean signature manifold of dimension $D=(d+1)$ with the $d$-dimensional boundary $Y$.
Near the boundary, we assume that the metric has a product structure so that in suitable local coordinates $(y^\alpha,\tau)$, we have $ds^2 = d\tau^2 + h_{\alpha\beta}(y) dy^\alpha dy^\beta$.
(Indices $\alpha,\beta,\ldots$ are tangent to $Y$.)

Let us now consider the Hilbert space $\hsp_{Y}$ associated with the Hamiltonian formulation of the theory in which $Y$ is regarded as a spatial manifold. 
Any choice of boundary condition $L$ on $Y$ determines a state $|L\> \in \hsp_Y$.
The path integral on $X$ for every choice of boundary condition on $Y$ similarly defines a state $|X\> \in \hsp_Y$.
Therefore the path integral computed with fixed boundary condition $L$ on $Y$ computes the overlap
\begin{align}
  Z(X, L) = \< L | X \>.
  \label{eq:ZX}
\end{align}
The proof essentially boils down to analyzing this path integral in two ways.  In one way, we analyze the effect of a symmetry transformation on the boundary state. In the other way, we analyze the effect of same symmetry transformation on the path integral measure à la Fujikawa.  Equating these two ways gives the \ac{APS} index theorem.  We will do this in \cref{sec:derivation}, but first we need to carefully look at boundary conditions and the associated boundary state.

\subsection{Boundary conditions and state correspondence}
\label{sec:bc-state}

To write down an index theorem on a manifold with boundary, we need to choose sensible boundary conditions which would allow us to define a bulk index.
We would like the boundary condition to be local. However, in many cases of interest, it turns out to be impossible to choose a sensible boundary condition which is also local.
Atiyah-Patodi-Singer \cite{atiyah_spectral_1975} discovered that the way to generalize the \ac{AS} index theorem is to give up on locality of the boundary condition. To describe the boundary condition, we note first note, near the boundary, we can treat the orthogonal coordinate $\tau$ as (Euclidean) time and take $\Gamma^\tau = \gamma^\tau$.
Thus, the \ac{KD} operator may be written as
\begin{align}
  \cK_X
  &= i\gamma^\tau (\del_\tau + \gamma^\tau \Gamma^\alpha D_\alpha) \nonumber\\
  &= i\gamma^\tau (\del_\tau + \cK_Y ).
    \label{eq:KY}
\end{align}
Here, $\cK_Y = \gamma^\tau \Gamma^\alpha D_\alpha$ is a Hermitian differential operator on the boundary manifold $Y$, which can be identified as the Hamiltonian operator on $\mathcal{H}_Y$.
With these definitions, the \ac{APS} boundary condition is
\begin{align}
	L \Phi \big|_{Y} &= + \Phi\big|_{Y},\quad \text{with}\quad
  L = \varepsilon(-\Ky), \label{eq:aps-bc}
\end{align}
where $\varepsilon(X) = X/\sqrt{X^{\dagger} X}$ is the sign function. In other words, the \ac{APS} boundary condition kills all the positive energy modes of $\Ky$.
The choice is self-adjoint, symmetric, elliptic although clearly nonlocal along the boundary. (See \cref{sec:boundary-conditions} for a discussion of these properties.)

\iffalse
The \ac{KD} operator $\cK_X$ is manifestly self-adjoint on manifolds without boundary.
For it to be self-adjoint on manifolds with boundary, however, we require the vanishing of the boundary term
\begin{align}
    \Delta &= \int_X d^D x\sqrt{g} \tr \Bigl[ \Psi^\dag (\cK_X \Phi) - (\cK_X \Psi)^\dag \Phi \Bigr] \nonumber\\
    &= \int_Y d^d y\sqrt{h} \tr \Psi^\dag \gamma^\tau \Phi .
    \label{eq:boundary-term}
\end{align}

It is convenient to decompose $\Phi$ into eigenvectors $\Phi_{\pm}$ of $\gamma^\tau$ where
\begin{align}
\gamma^\tau \Phi_\pm = \pm \Phi_\pm,
\end{align}
and similarly for $\Psi$.
This lets us rewrite the boundary term as
\begin{align}
  \Delta = \int_Y d^{d} y \sqrt{h} 
  \Bigl[ \tr \Psi^\dag_+ \Phi_+ -\tr \Psi^\dag_- \Phi_- \Bigr]
\end{align}
We see that a general choice of boundary condition that makes this vanish is 
\begin{align}
  &\Phi_+ = U (\Phi_-)
\end{align}
where $U$ is a isometric transformation from the $\gamma_{\tau} = -1$ space to the $\gamma_\tau = +1$ space.
For example, in the absence of any zero modes of $K$, one can make the choice $U = K_{-+}/\sqrt{ K_{-+}K_{+-} }$. 
These are called the standard \ac{APS} boundary conditions.\footnote{This is true when the boundary theory has no zero modes. In the presence of zero-modes, the definition of \ac{APS} boundary conditions becomes more subtle, as described in Ref.~\cite{witten_anomaly_2020}.}
(Note that these are nonlocal boundary conditions. For more discussion on local boundary conditions, see the next section.)
\fi

What is the boundary state corresponding to the \ac{APS} boundary conditions? Actually, these boundary conditions have a very natural physical interpretation, as shown in Ref.~\cite{yonekura_daifreed_2016}.
The free-fermion Hamiltonian for the theory on the spatial manifold $Y$ can be written as
$ H = \sum_{\lambda, \sigma} \lambda c_{\lambda,\sigma}^\dagger c_{\lambda, \sigma}^{\xdagger} , $
where the sum is over the eigenvalues $\lambda$ of $\cK_{Y}$  defined in \cref{eq:KY}, and $\sigma$ represents other discrete labels.
The \ac{APS} boundary conditions in \cref{eq:aps-bc} kill all the positive energy modes of $\Ky$. Under the correspondence derived by Ref.~\cite{yonekura_daifreed_2016}, this corresponds to the state in the Hilbert space with all the positive energy modes empty, while all the negative energy modes are filled:
\begin{align}
	c_{\lambda, \sigma}^{\xdagger} | \Omega \> &= 0 \quad \lambda > 0, \\
	c_{\lambda, \sigma}^{\dagger} | \Omega \> &= 0 \quad \lambda < 0.
\end{align}
But this is precisely the ground state $|\Omega\>$ of the massless fermion Hamiltonian, which is characterized by filling the Dirac sea. Therefore the boundary state corresponding to the \ac{APS} boundary conditions is the massless free-fermion vacuum.

The other important ingredient in the \ac{APS} index theorem is the $\eta$-invariant of a boundary Dirac operator. In this picture, the $\eta$ has an elegant physical interpretation as well. It is the chiral charge of the massless vacuum state. This is true for both notions of the chiral symmetry.

To see this, assume that there is an operator $\GammaChi$ which satisfies $\GammaChi^{2}=1$ and $[\GammaChi, \Ky]=0$.
We can simultaneously diagonalize $\GammaChi$ and $\Ky$ to define the eigenmodes $\Phi_{\lambda, \pm}$ which satisfy
\begin{align}
  \Ky \Phi_{\lambda, \pm} = \lambda \Phi_{\lambda, \pm}, \quad\quad
  \GammaChi \Phi_{\lambda, \pm} = \pm \Phi_{\lambda, \pm},
\end{align}
which means that the corresponding particle creation/annihilation operators can be written as $c_{\lambda,\pm}$.
With the number operators
$N_{\lambda, \pm} = c_{\lambda, \pm}^\dagger c_{\lambda, \pm}^{\xdagger}$,
the charge operator for the $\GammaChi$ symmetry is
\begin{align}
 \QA = \sum_{\lambda} (N_{\lambda,+} - N_{\lambda,-}).
\end{align}
For \ac{KD} fermions with chiral symmetry $\GammaChi = \GammaGB$ or $\GammaH$, there is a pairing of $(\lambda, +)$ modes and $(-\lambda, -)$ modes.
This is because $\gamma^0$ anticommutes with both $\Ky$ and $\GammaChi$, there is a pairing of the eigenfunctions $\Phi_{\lambda, +}$ and $\gamma^0 \Phi_{\lambda, +}$ such that
$
  \gamma^0 \Phi_{\lambda, +} = \Phi_{-\lambda, -}.
$
Since the ground state is characterized by filling up all the negative $\lambda$ modes,
the operators $\sum_{{\lambda}} N_{\lambda, \pm}$ acting on $|\Omega \>$ simply counts the number of  $\pm$ chirality modes with $\lambda <0$. But the $\pm$ pairing under $\gamma^{0}$ implies that the number of $\lambda<0$ positive chirality modes is the same as the number of $\lambda > 0$ negative chirality modes. Therefore, the chiral charge of the ground state $ |\Omega \>$ is
\begin{align}
  \QA |\Omega \rangle
  &= \Bigl( \sum_{\lambda>0} 1 - \sum_{\lambda<0} 1 \Bigr) |\Omega \> \nonumber\\
  &= \eta(\Ky^{-}) |\Omega \>.
  \label{eq:charge-eta}
\end{align}
where $\Ky^{-}$ is the restriction of the boundary \ac{KD} operator $\Ky$ to the negative chirality subspace. We see that the chiral charge of the massless vacuum state $| \Omega \>$ is naturally given by the quantity $\eta(\Ky^{-})$ which is a measure of the ``spectral asymmetry'' of the operator $\Ky^{{-}}$. This is the $\eta$-invariant.

The above derivation is valid for both chiral symmetries. However, for the case of twisted chiral symmetry, there is an important simplification.  
If $\Phi_{\lambda, \pm}$ is an eigenfunction of $\Ky$ with $\GammaGB = \pm$, we note that the operator which sends $C_{\tau}\colon \Phi \to \gamma^{\tau} \Phi \gamma^{\tau}$ anticommutes with $\Ky$ but commutes with $\GammaGB$. Therefore $C_{\tau}(\Phi_{\lambda, \pm}) = \Phi_{-\lambda, \pm}$ and this exact pairing of positive and negative modes of the same chirality causes the $\eta$ invariant to vanish.

\subsection{Derivation of the APS index theorem}
\label{sec:derivation}

In this section, we derive the \ac{APS} theorem for the \ac{KD} operator.
This will lead to the Chern--Gauss--Bonnet and Hirzebruch signature theorems on manifolds with boundary.

We have seen that on a closed manifold the local change of variables
\begin{align}
    \Psi'(x) = e^{i\alpha(x)\GammaChi} \Psi(x), \quad \Psibar'(x) = \Psibar(x) e^{i\alpha(x)\GammaChi},
    \label{eq:chiral-transf}
\end{align}
gives rise to a Jacobian factor in the fermion measure:
\begin{align}
    [d\Psibar'][d\Psi'] = [d\Psibar][d\Psi] e^{-2i\int_X \alpha G},
    \label{eq:Jacobian}
\end{align}
where for twisted chiral symmetry $G$ is the Euler density while for untwisted chiral symmetry it is the $L$ polynomial. 
By locality, this formula for the Jacobian should still be good on a manifold with boundary provided that $\alpha$ vanish near the boundary.
On the other hand, if $\alpha(x)$ is equal to a constant $\alpha_0$, then the Jacobian takes the form
\begin{align}
    [d\Psibar'][d\Psi'] = [d\Psibar][d\Psi] e^{-2i\alpha_0 I},
\end{align}
where $I$ is the appropriate index of the \ac{KD} operator
\begin{align}
    I = \tr \GammaChi \mathcal{K}.
\end{align}
Thus, for twisted chiral symmetry $I$ is the Euler characteristic and for untwisted chiral symmetry it is the signature.
The same argument goes through even in the presence of a boundary, as long as the boundary conditions are such that $\mathcal{K}$ is self-adjoint.
In particular, the path integral vanishes unless we insert an operator $\mathcal{O}$ with chiral charge $2 \ind$.

The proof of the \ac{APS} theorem will consist in comparing two expressions for the expectation value 
\begin{align}
    \langle e^{i \alpha_0 Q} \mathcal{O}(x_0) \rangle_X
    = \langle \Omega| e^{i \alpha_0 Q} |X, \mathcal{O}\rangle
    \label{eq:corr-func}
\end{align}
where $Q$ is the chiral charge operator on the Hilbert space for the $\GammaChi$ symmetry.
First, we can let $e^{i\alpha_0 Q}$ act to the left on $\langle \Omega|$, so that
\begin{align}
    \langle \Omega | e^{i \alpha_0 Q} | X,\mathcal{O} \rangle
    = e^{i\alpha_0 \eta }\langle \Omega|X,\mathcal{O}\rangle
    \label{eq:aps-lhs}
\end{align}
where we used the chiral charge of the vacuum given by \cref{eq:charge-eta} in terms of the eta invariant, $\eta  \equiv \eta(\Ky^{-})$.

\begin{figure}
    \centering
    \includegraphics[width=0.9\linewidth]{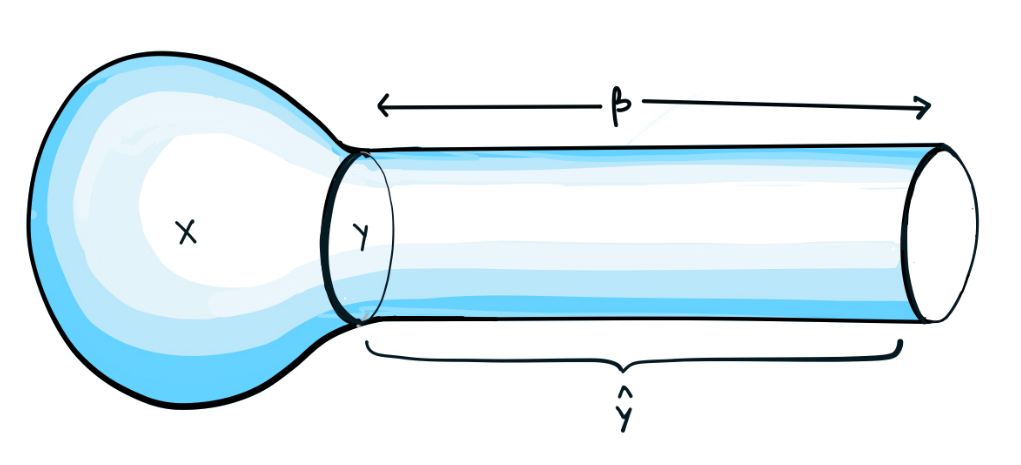}
    \caption{The extended manifold $\hat{X}$ obtained by attaching the cylinder $\hat Y = Y \times [0, \beta]$ to the boundary $Y$ of original manifold $X$.}
    \label{fig:Xhat-manifold}
\end{figure}

On the other hand,  we can interpret the correlation function in \cref{eq:corr-func} as a path integral on a manifold $\hat {X}$ defined as follows.
We first write the expectation value as 
\begin{align}
    \< \Omega | e^{i \alpha_0 Q} | X, \mathcal{O} \> = \< \Omega | e^{-\beta H } e^{i \alpha_0 Q} | X, \mathcal{O} \>.
\end{align}
where we have assumed that the ground state energy vanishes for simplicity.
The expression on the right hand side can be interpreted as a path integral on a manifold $\hat{X}$ which is defined by attaching a cylinder $Y \times [0,\beta]$ with the product metric to the boundary $Y$ of $X$, as shown in \cref{fig:Xhat-manifold}.
Thus,
\begin{align}
\langle \Omega | e^{-\beta H} e^{i \alpha_0 \int_Y J } |X,\mathcal{O}\rangle &=
    \< e^{i\alpha_0 \int_Y J} \mathcal{O}(x_0) \>_{\hat{X}} \nonumber\\
    &= \int [d\Psibar][d\Psi] e^{-S_{\hat{X}}} e^{i\alpha_0 \int_Y J}  \mathcal{O}(x_0),
    \label{eq:aps-almost-rhs}
\end{align}
where the fields are taken to live on $\hat{X}$ and $J$ is the chiral symmetry current.
If we now perform the change of variables given in \cref{eq:chiral-transf} with $\alpha(x)$ equal to $\alpha_0$ on $X$ and $0$ on $\Hat{Y}$, then according to \cref{eq:Jacobian}, the change in the measure is given by
\begin{align}
    [d\Psibar][d\Psi] \to [d\Psibar][d\Psi] e^{- 2 i \int_{\hat X} \alpha G} = [d\Psibar][d\Psi]e^{-2 i \alpha_0 \int_{X} G}.
\end{align}
We also have a change in the action given by
\begin{align}
    S \to S + i \int_{\hat{X}} d \alpha \wedge J  = S + i \alpha_0 \int_Y J,
\end{align}
where the last expression comes from the fact that $d\alpha$ is a delta function 1-form supported on $Y$. 
Finally, we have a change in $\mathcal{O}(x_0)$ given by
\begin{align}
    \mathcal{O}(x_0) \to e^{2 i \alpha_0 I } \mathcal{O}(x_0).
\end{align}
So with this change of variables, the right-hand side of \cref{eq:aps-almost-rhs} can be rewritten as
\begin{align}
    &\langle \Omega | e^{-\beta H } e^{i\alpha_0 \int_Y J}|X,\mathcal{O}\rangle \nonumber\\
    &\quad= \int [d\Psibar][d\Psi] e^{-S} \mathcal{O}(x_0)\ e^{ 2 i \alpha_0 I  - 2 i\alpha_0 \int_X G } \nonumber\\
    &\quad= \langle \Omega |X,\mathcal{O}\rangle\ e^{ 2 i \alpha_0 I  -2 i\alpha_0 \int_X G }.
    \label{eq:aps-rhs}
\end{align}
Comparing \cref{eq:aps-lhs,eq:aps-rhs}, we finally obtain
\begin{align}
    I  = \int_X G + \frac12 \eta,
\end{align}
which is precisely the \ac{APS} index theorem for \ac{KD} fermions.

At this point, we can now comment on the \ac{APS} index theorems for the two chiral symmetries. For the untwisted chiral symmetry, we obtain the Hirzebruch signature theorem on manifolds with boundary: $I$ is the signature, $G$ is the Hirzebruch $L$-polynomial, and the $\eta $ invariant gives a contribution from the boundary. On the other hand, in the case of the twisted chiral symmetry $\GammaGB$, the index is the Euler characteristic and $G$ is the Euler density. However, as noted at the end of \cref{sec:bc-state},  the twisted chiral charge of the massless vacuum vanishes due the pairing of positive and negative eigenmodes. Therefore, the \ac{APS} index theorem in this case is just the \ac{GB} theorem $\chi(X) = \int e(X) $, with $e(X)$ being the Euler density of the manifold $X$. In the usual presentation of the \ac{GB} theorem, there is an additional contribution from the geodesic curvature of $Y$ embedded in $X$. However, the requirement of product structure near the boundary implies that the geodesic curvature vanishes.

\section{Local, symmetric boundary conditions}
\label{sec:boundary-conditions}

The fact that there is no boundary contribution for the \ac{GB} theorem was noted in the motivation of the original work of \ac{APS} \cite{atiyah_spectral_1975}.  For the twisted chiral symmetry, the obstruction to finding suitable symmetric, local, self-adjoint boundary conditions for the \ac{KD} operator vanishes, while for the untwisted chiral symmetry, it does not.

To gain some understanding of the difference between to the two cases, let us consider for simplicity a local boundary condition of the form
\begin{align}
  L(\Phi|_Y) = A (\Phi|_{Y}) B = +\Phi|_{Y}.
  \label{eq:L-AB}
\end{align}
for some matrices $A$ and $B$. 
In general, we require that $L^2 = 1$ (and hence $A^2=B^2=1$), and that the $L=+1$ and $L=-1$ eigenspaces have the same dimensions so that the boundary condition eliminates half of the modes.
Another general requirement is ellipticity, which we will comment on below.
But first, let us address what constraints on $A,B$ are needed in order for the \ac{KD} operator to be self-adjoint or symmetric. 

\subsection*{Self-adjointness and symmetries}

In order to have $\cK$ self-adjoint, the boundary condition must force the vanishing of the boundary term
\begin{align}
 \Delta
  &= \int_Y d^dy\sqrt{h} \tr [\Psi^\dag \gamma^{\tau} \Phi] \nonumber\\
 &= \int_Y d^dy\sqrt{h} \tr [L(\Psi)^\dagger \gamma^\tau L(\Phi) ].
   \label{eq:bdy-hermitian}
\end{align}
We see that it suffices to take $L$ to be unitary and to anticommute with $\gamma^\tau$.
In the case of the local boundary condition \eqref{eq:L-AB}, this means
\begin{align}
  &A A^\dag = B B^\dag = 1, \label{eq:hermitian-condition1}\\
  &A^\dag \gamma^\tau A = - \gamma^\tau.
  \label{eq:hermitian-condition2}
\end{align}

For the boundary condition to preserve a symmetry, $L$ must commute with the symmetry transformation.
Let us see how this may be achieved for the local boundary condition \eqref{eq:L-AB}.
Consider first rotation invariance along the boundary.
For this, we require $A,B$ to commute with the rotation generators $\sigma_{ab}$ with $a,b$ along the boundary. 
The most general choice of $A$ or $B$ which commutes with the $\sigma_{ab}$ are linear combinations of the form
\begin{align}
  A,B  = a +  b \gamma^{\tau} + c \gammabar + d i\gammabar\gamma^{\tau}.
  \label{eq:AB-rotational-invariance}
\end{align}

Next, let us look at the two chiral symmetries.
To preserve untwisted-chiral symmetry, we need $[A,\gammabar] = 0$, while $B$ can be arbitrary.  We only have the choices $A=\Id, \gammabar$.

To preserve twisted-chiral symmetry, we need both $A,B$ to either commute or anticommute with $\gammabar$.
For the commuting case, we can have
    \begin{align}
        (A,B) = (\gammabar,1) , \, (\gammabar,\gammabar), \, (1,\gammabar),
        \label{eq:chiral-bc}
    \end{align}
which will also preserve untwisted-chiral symmetry. 
For the anticommuting case, we can have
    \begin{align}
        A, B &= i \gammabar e^{{i \theta \gammabar}} \gamma^{\tau},
        \label{eq:complex-chiral-bag}
    \end{align}
where $\theta{}$ may be different for $A$ and $B$.
Note however that this will not preserve untwisted chiral symmetry.
The analog of this for Dirac fermions (where $B$ plays no role) has been called
``Euclidean (complex) chiral-bag'' boundary conditions
\cite{hrasko_fermion_1984, wipf_gauge_1995, chodos_baryon_1974, beneventano_strong_2003, chodos_new_1974}, which also does not preserve chiral symmetry.
We shall refer to the boundary condition of \cref{eq:complex-chiral-bag} as the ``twisted'' chiral-bag boundary condition.

Comparing with \cref{eq:hermitian-condition1,eq:hermitian-condition2}, we note that
the twisted chiral-bag boundary condition is only self-adjoint for $\theta=0$.
The case of $\theta = \pm \pi/2$, while not self-adjoint, is also of interest. 
The analog of it for Dirac fermions was considered in Ref.~\cite{witten_anomaly_2020}, so we call it the twisted \ac{WY} boundary condition.

At this point, it seems that a choice such as $A = \gammabar$, $ B= 1$ consistent with \cref{eq:chiral-bc} has the virtues of ensuring self-adjointness of the Dirac operator [since it satisfies \cref{eq:hermitian-condition1,eq:hermitian-condition2}] and of preserving both chiral symmetries.
But it turns out to be unacceptable for a different reason -- it is not elliptic.

\subsection*{Ellipticity}

In order for a boundary condition to even be sensible, it must be elliptic.
This means the following.
We take $X = Y \times \RR^{+}$ with $Y = \RR^{d}$.
We ask whether there are any normalizable zero-modes of $\Kx = i \gamma^{\mu} \del_{\mu}$ satisfying the boundary condition $L$.
If no solution exists, then we say the boundary condition is (strongly) elliptic.

Let us now determine when the boundary condition is elliptic.
With the ansatz $\Psi(x, \tau) = \Phi(x) e^{- \lambda \tau}$,
the equation for a zero mode $\Kx \Psi = 0 $ becomes
\begin{align}
	 \Ky \Phi = \lambda \Phi \quad (\lambda > 0).
  \label{eq:ellipticity-cond}
\end{align}
Now, if the operator $L$ commutes with $\Ky$, then $\Ky$ and $L$ can be simultaneously diagonalized.
Moreover, if $L$ is local then for each $\lambda > 0$ in the spectrum of $\Ky$, there is a normalizable zero mode of $\Kx$ with $L = +1$.
In the continuum limit, this means that there are infinitely many zero modes localized to the wall with arbitrarily high momenta along the boundary.
This does not lead to a well-defined $(d+1)$-dimensional Euclidean field theory.

On the other hand, if $L$ anticommutes with $\Ky$, then $L\Phi$ is an eigenstate of $\Ky$ with eigenvalue $-\lambda$.
Since $\lambda$ is nonzero, it follows that $L\Phi \neq \Phi$.
Therefore, there is no solution with $\lambda > 0$ which satisfies the boundary condition $L$. 
Such a boundary condition is therefore elliptic.%
\footnote{It is interesting to note that the ellipticity of a local boundary condition here also ensures the Hermiticity of the Hamiltonian on a spatial manifold with boundary. 
To see this, let $Y$ have coordinates $t,\Vec{y}$, and regard the $d$-dimensional surfaces of constant $t$ as spatial slices, so that the Hamiltonian is given by $H = i \gamma^t (\Vec{\gamma} \cdot \Vec{\del} + \gamma^\tau \del_\tau)$.
In order for this Hamiltonian to be Hermitian, we need the vanishing of the boundary term
$\int d^{d-1} y \ \tr \Psi^\dag \gamma^t \gamma^\tau \Phi|_{\tau=0,t=\text{const.}}$.
Thus, if $L$ is elliptic and local, then $L$ anticommutes with $\gamma^\tau \gamma^a$ for all $a$ tangent to $Y$ and in particular with $\gamma^\tau \gamma^t$, so that this boundary term vanishes.
}

Clearly, the boundary condition in \cref{eq:chiral-bc} commutes with $\Ky$ and is therefore not elliptic.  On the other hand, the (twisted) chiral-bag boundary condition given in \cref{eq:complex-chiral-bag} anticommutes with $\Ky$ and is therefore elliptic.  Since ellipticity is a necessary condition for the Euclidean field theory to be well-defined \cite{witten_anomaly_2020}, the twisted chiral-bag boundary condition is the only sensible choice.
(See also Refs.~\cite{beneventano_strong_2003, ivanov_anomaly_2022} for discussions of ellipticity of the chiral-bag boundary conditions.)

In the usual presentation for Dirac fermions, the chiral-bag boundary condition has been considered
\cite{chodos_new_1974, chodos_baryon_1974, kurkov_parity_2017a, ivanov_anomaly_2022, kurkov_gravitational_2018, fialkovsky_quantum_2019}.
But in that case, as for the untwisted chiral symmetry for \ac{KD} fermions, it is local, self-adjoint, and elliptic, but breaks the bulk chiral symmetry.
The surprising fact here is that it has a twisted version, which preserves the twisted chiral symmetry.
This lets us define a bulk index even in the presence of a boundary.
This is completely consistent with the fact that the Euler characteristic can be defined for manifolds with a boundary.
In fact, for the untwisted chiral symmetry there is no such local boundary condition, which necessitates the use of the nonlocal \ac{APS} boundary conditions.

With the twisted chiral-bag boundary condition, one can proceed with the derivation of \ac{APS} index theorem as in the previous section.
Since the boundary-condition is local and symmetric, the twisted chiral charge of the boundary state vanishes, and there is no contribution from the boundary  $\eta$ invariant.
This is another way to understand the result derived earlier about there being no boundary contribution to the \ac{GB} theorem.

Before closing, we note that the nonlocal \ac{APS} boundary conditions can be easily seen to be self-adjoint, symmetric and elliptic.   Using the expression $L = -\varepsilon(\Ky)$ from \cref{eq:aps-bc}, it is clear that $L$ is unitary and anticommutes with $\gamma^{\tau}$, and is therefore self-adjoint [see \cref{eq:bdy-hermitian}].  It also commutes with $\gammabar$, so preserves (both notions of) bulk chirality.  Finally, ellipticity: as discussed earlier, a normalizable zero-mode of $\Kx$ corresponds to a solution of \cref{eq:ellipticity-cond} with $\lambda > 0$. But these are exactly the modes killed by the \ac{APS} boundary condition $L = +1$. Therefore, the \ac{APS} boundary condition is elliptic.

\section{Conclusion}%
\label{sec:conclusion}

Recently, \ac{KD} fermions have found applications in attempts to use \ac{SMG} as an ingredient of chiral lattice gauge theories, as well as in explaining the classification of bosonic \ac{SPT} phases.
This is because \ac{KD} fermions have a natural lattice discretization which preserves many features of the continnum, including some 't~Hooft anomalies.
On the other hand, the \ac{APS} index theorem has been recognized as an important ingredient in obtaining a nonperturbative picture of anomaly cancellation for fermionic theories.
Motivated by these connections, in this work, we derive the \ac{APS} index theorem using a ``physics'' approach due to Ref.~\cite{kobayashi_atiyahpatodisinger_2021}, which generalizes Fujikawa's method to manifolds with boundary.  
In this approach, the connection to anomaly inflow is manifest -- the boundary state may be charged under the chiral symmetry, and this causes an additional contribution to the usual \ac{AS} index theorem from the boundary. 
This extra contribution is precisely the $\eta$ invariant of the boundary Dirac operator.

An interesting feature of \ac{KD} fermions, which distinguishes them from Dirac fermions, is that there are two notions of a chiral symmetry. 
In flat space, where \ac{KD} fermions can be decomposed into $2^{{D/2}}$ Dirac flavors, one of them corresponds to the usual chiral symmetry $\GammaH$ and other symmetry $\GammaGB$ corresponds to a mixed chiral--flavor transformation, hence the name ``twisted'' chiral symmetry.
Both of these have a perturbative mixed anomaly with gravity, and therefore lead to index theorems via Fujikawa's method.
It is known that (for example, see Ref.~ \cite{green_superstring_2012}) the index theorems for the twisted- and untwisted-chiral symmetries correspond to the Chern-Gauss-Bonnet and Hirzebruch signature theorems, respectively.

When trying to generalize the \ac{AS} index theorem to manifolds with boundary, it is important to choose the boundary conditions carefully.
It was noted by \ac{APS} \cite{atiyah_spectral_1975} that the attempt to extend Hirzebruch signature theorem to manifolds with boundary meets with an obstruction to finding sensible local boundary conditions.
Therefore, in trying to define an index, one is forced to consider the nonlocal \ac{APS} boundary conditions.  
These boundary conditions have a natural interpretation as those corresponding to a massless fermion ground state at the boundary \cite{yonekura_daifreed_2016}, and the $\eta$ invariant appears as the chiral charge of this boundary state for the Hirzebruch theorem.  
On the other hand, for the \ac{GB} theorem, we find that the boundary charge vanishes for the \ac{APS} state. 
This is why there is no contribution from the boundary $\eta$ invariant for the index theorem associated with the twisted chiral symmetry.

The fact that there is no boundary correction to the \ac{GB} theorem can also be understood in another way.  
If we can find a ``good'' local and symmetric boundary, which corresponds to some state in the boundary fermion Hilbert space, then we could use that instead of the \ac{APS} state in our proof, and the charge would trivially vanish.  
We found that such a boundary condition can in fact be defined. 
In the context of Dirac fermions, these boundary conditions have been called Euclidean chiral-bag boundary conditions. 
However, an important difference from the case of Dirac fermions (and untwisted chiral symmetry for \ac{KD} fermions) is that the chiral-bag boundary conditions are not chirally symmetric in that case, while the twisted boundary conditions are.
The existence of such local, symmetric boundary conditions for twisted chiral symmetry is consistent with the fact that the Euler characteristic can be defined for manifolds with boundary with no problem.

Recently, Catterall \cite{catterall_lattice_2023} has argued that the problem of chiral fermions has an analog for \ac{KD} fermions.
Using the twisted-chiral symmetry allows one to solve this problem.
This is due to the ``onsite'' nature of the twisted-chiral symmetry on the lattice, which plays an important role in demonstrating anomaly cancellation on the lattice.
In line with this, there has been a significant improvement in the understanding of how anomalies can arise on the lattice with onsite symmetries
\cite{nguyen_lattice_2023a, catterall_hooft_2022, cheng_liebschultzmattis_2023, fazza_lattice_2023, gorantla_modified_2021a, seifnashri_liebschultzmattis_2024}.
This onsite nature of the twisted symmetry makes the analogous problem of chiral lattice gauge theory much more tractable \cite{catterall_lattice_2023} than L\"uscher's approach where one needs to define a nonlocal $\gamma_{5}$ matrix using \ac{GW} fermions \cite{Luscher:1998pqa, luscher2001chiral, luscher1999abelian, poppitz_chiral_2010}.
In the genuinely difficult case of (untwisted) chiral symmetry where there is no known onsite lattice discretization, one has to gauge non-onsite symmetries
\cite{luscher1999abelian, catterall_gauging_2024, seifnashri_liebschultzmattis_2024}.
As we have argued, the twisted-chiral symmetry is also special from another perspetive in that there is no obstruction to formulating a symmetric, local boundary condition, and therefore the \ac{APS} index theorem has no boundary contribution from the $\eta$ invariant.
This might provide another way to understand the role played by twisted-chiral symmetry in this problem. This is also interesting in light of Kaplan's recent proposal \cite{kaplan_weyl_2024, kaplan_chiral_2024} with a single domain-wall where the role of the \ac{APS} index theorem is manifest.
A lattice version of the \ac{APS} index theorem may help tie the various approaches together.
(For example, see also Refs.~\cite{fukaya_modtwo_2022, fukaya2020atiyah} for an approach to formulating the \ac{APS} index theorem on the lattice which uses domain-wall fermions.)
This is likely to be important in understanding the problem of formulating chiral gauge theory on the lattice.

\section*{Acknowledgements}

We thank Simon Catterall and David Kaplan for inspiring conversations.
M.N. is supported by the U.S. Department of Energy, Office of Science, Office of Nuclear Physics under Award Number DE-FG02-03ER41260.
H.S. is supported
by the Department of Energy through the Fermilab QuantiSED program in the area of ``Intersections of QIS and Theoretical Particle Physics."
Fermilab is operated by Fermi Research Alliance, LLC under Contract No. DE-AC02-07CH11359 with the United States Department of Energy.

\bibliography{refs}

\appendix

\section{Review of differential forms}
\label{sec:dforms}

Differential $p$-forms are totally antisymmetric covariant tensor fields with $p$ indices, $\phi_{\mu_1\ldots\mu_p}$.
A coordinate invariant way to express these objects is to introduce totally antisymmetric products of coordinate differentials $dx^{\mu_1} \wedge \cdots \wedge dx^{\mu_p}$ and write
\begin{align}
    \phi^{(p)} = \frac{1}{p!} \phi_{\mu_1\ldots\mu_p} dx^{\mu_1} \wedge \cdots \wedge dx^{\mu_p},
\end{align}
The product of differential forms $\phi^{(p)}\wedge \psi^{(q)}$ is then defined in the obvious way.

A natural derivative of differential forms is the exterior derivative $d$, defined by
\begin{align}
    d \phi^{(p)} = dx^\mu \wedge \del_\mu \phi^{(p)}. 
\end{align}
This derivative is clearly coordinate invariant and independent of the metric. 
A crucial property is nilpotency: $d^2 = 0$. 

Another important operation on differential forms, available when the manifold has a metric $g_{\mu\nu}$, is the Hodge star operation $\star$ which transforms a $p$-form into a $(D-p)$-form. 
In components, it is defined by
\begin{align}
    (\star \phi)_{\mu_{p+1}\ldots\mu_D} 
    = \frac{1}{p!} \sqrt{g} \epsilon_{\mu_1\ldots \mu_p \mu_{p+1}\ldots\mu_D} \phi^{\mu_1 \ldots \mu_p},
\end{align}
where $\epsilon_{\mu_1\ldots\mu_D}$ is totally antisymmetric Levi-Cevita symbol.
It is straightforward to show that
\begin{align}
    \star^2 = (-1)^{p(D-p)}
\end{align}
on $p$-forms.

With the help of the Hodge star, an inner product on the space of $p$-forms can be defined by
\begin{align}
    (\psi^{(p)},\phi^{(p)}) 
    &= \int \psi^{(p)*} \wedge {\star \phi^{(p)}} \nonumber\\
    &= \frac{1}{p!} \int \psi^*_{\mu_1\ldots\mu_p} \phi^{\mu_1\ldots\mu_p} \sqrt{g}d^D x,
\end{align}
where in the last expression indices are raised with the metric tensor.

Given this inner product structure, there is another natural derivative $\delta$, defined as the formal adjoint of $d$:
\begin{align}
    (\psi^{(p)}, d \phi^{(p-1)}) = (\delta \psi^{(p)}, \phi^{(p-1)}).
\end{align}
It is straightforward to express $\delta$ in terms of the exterior derivative and the Hodge star:
\begin{align}
    \delta = (-1)^p \star^{-1} d \star
\end{align}
on $p$-forms.
Note that the nilpotency of $d$ also implies the nilpotency of $\delta$.

The Laplacian of arbitrary differential forms is given by
\begin{align}
    \Delta = d \delta + \delta d.
\end{align}
The K\"{a}hler-Dirac operator is a square-root of this Laplacian:
\begin{align}
    \cK = i(d-\delta).
\end{align}
Note that it does not require a spin structure to be defined. 

\section{Review of vielbeins}
\label{sec:vielbein}

An orthonormal frame (or ``vielbein'') is a collection of $D$ orthonormal vector fields $e_{m\mu}$ ($m=1,\ldots,D$):
\begin{align}
    g^{\mu\nu} e_{m\mu} e_{n\nu} = \delta_{mn}.
\end{align}
The orthonormality of the frame also implies the completeness relation
\begin{align}
    e_{m\mu} e_{m\nu} = g_{\mu\nu}. 
\end{align}
(Summation over repeated Latin indices is understood.)
Here, we are adopting an index convention where Greek indices $\mu,\nu,\ldots$ are ``curved space'' indices, while Latin indices $m,n,\ldots$ are ``flat space'' indices, which simply label the various vectors in the frame.

If we have matrices $\gamma_m$ that satisfy the flat space Clifford algebra
\begin{align}
    \{\gamma_m,\gamma_n\} = 2 \delta_{mn},
\end{align}
then the matrices $\Gamma_\mu \equiv \gamma_m e_{m\mu}$ will satisfy the curved space Clifford algebra
\begin{align}
    \{\Gamma_\mu,\Gamma_\nu\} = 2 g_{\mu\nu}.
\end{align}
Under a frame rotation
\begin{align}
    e_{m \mu} \to e_{n \mu} R_{nm},
\end{align}
where $R_{mn}$ is a matrix in $SO(D)$, the curved space gamma matrices transform as
\begin{align}
    \Gamma_\mu \to S^{-1}(R) \Gamma_\mu S(R),
\end{align}
where $S(R)$ is a lift of $R$ to the spin group $\Spin(D)$.
The matrix representation of the \ac{KD} field, $\Phi$, transforms in the same way under such a frame rotation:
\begin{align}
    \Phi \to S^{-1}(R) \Phi S(R).
\end{align}
Note that the sign ambiguity in choosing the lift $S(R)$ cancels out in these transformation laws.
This of course reflects the fact that no spin structure is required to define \ac{KD} fields on arbitrary manifolds.

We can define a covariant derivative for \ac{KD} fields by
\begin{align}
    D_\mu \Phi = \del_\mu \Phi + \tfrac{1}{2}\omega_{mn \mu}[\sigma_{mn},\Phi].
\end{align}
Here, $\omega_{mn\mu}$ is the spin connection, defined by
\begin{align}
    \nabla_\mu e_{n\nu} = e_{m\nu} \omega_{mn \mu},
\end{align}
where $\nabla_\mu$ is the ordinary Levi-Cevita derivative operator; and the $\sigma_{mn}$ are the generators of $\Spin(D)$:
\begin{align}
    \sigma_{mn} = \tfrac{1}{4}[\gamma_m,\gamma_n].
\end{align}
The \ac{KD} operator can then be written as $\cK = i \Gamma^\mu D_\mu$. 

With the help of the product rule
\begin{align}
    &\nabla_\mu \tr \Psi^\dag \Gamma^\mu \Phi = \tr [(D_\mu \Psi)^\dag \Gamma^\mu \Phi + \Psi^\dag \Gamma^\mu D_\mu \Phi],
\end{align}
we see that on closed manifolds, Hermiticity of the \ac{KD} operator is manifest, while on manifolds with boundary, Hermiticity requires the vanishing of the boundary term
\begin{align}
    \Delta = \int_Y d^d y \sqrt{h} \tr \Psi^\dag \gamma^\tau \Phi.
\end{align}
Boundary conditions that ensure $\Delta=0$ are discussed in the main text. 

\section{Untwisted chiral symmetry}
\label{sec:ChiralHsym}
In this appendix, we derive the form of the chirality operator $\GammaChi = \GammaH$ that exchanges $d$ and $\delta$. 

The formula $\delta = (-1)^k \star^{-1} d \star$, where $\star$ is the Hodge star operation on forms (Appendix \ref{sec:dforms}), suggests the ansatz $\GammaH = \star B$ where $B$ acts as degree-dependent phase factor.
Writing $B=B_k$ for the action of $B$ on $k$-forms, the requirement that $\GammaH$ exchange $d$ and $\delta$ gives
\begin{align}
  \delta 
  &= \GammaH^{-1} d \GammaH \nonumber\\
  &=B^{-1} (\star^{-1} d \star) B \nonumber\\
  &= (-1)^{k} B^{-1} \delta B \nonumber\\
  &= (-1)^{k} \frac{B_{k}}{B_{k-1}} \delta.
\end{align}
This  gives the recursion relation $B_{k} = (-1)^{k} B_{k-1}$,
which can be solved to give
$B_{k} = (-1)^{g_{k}} B_0$, where $g_k = \frac12 k(k+1)$.
On the other hand, the requirement that $\GammaH$ square to the identity gives
\begin{align}
  1
  &= \GammaH^2 \nonumber\\
  &= B_{D-k}B_{k} (-1)^{k(D-k)} \nonumber\\
  &= (B_0)^2(-1)^{g_{D-k} + g_{k} + k(D-k)}  \nonumber\\
  &= (B_0)^2 (-1)^{g_D}.
\end{align}
where in the second line we have used the fact $\star^2 = (-1)^{k(D-k)}$ on $k$-forms. 
Therefore, we can choose $B_0 = (\pm i)^{g_D}$, which finally gives
\begin{align}
  B_{k} = (-1)^{g_k} (\pm i)^{g_D}.
\end{align}
The overall sign $(\pm 1)^{g_D}$ may be chosen arbitrarily.
The operator $B$ is referred to as the main anti-automorphism in Ref.~\cite{becher_dirackahler_1982}.

\end{document}